# Simultaneous tensile and shear measurement of the human cornea *in vivo* using S0- and A0-wave optical coherence elastography

Guo-Yang Li[1,†,#], Xu Feng[1,†], Seok-Hyun Yun[1,2,*]

[1] Harvard Medical School and Wellman Center for Photomedicine, Massachusetts General Hospital, 50 Blossom St., Boston, MA 02114, USA.

[2] Harvard-MIT Division of Health Sciences and Technology, Cambridge, MA 02139, USA.

[#] Current address: Department of Mechanical Engineering, Peking University, Peking, China

[†] Equal contribution

[*] Correspondence: syun@hms.harvard.edu

**Abstract:** Understanding corneal stiffness is valuable for improving refractive surgery, detecting corneal abnormalities, and assessing intraocular pressure. However, accurately measuring the elastic properties, particularly the tensile and shear moduli that govern mechanical deformation, has been challenging. To tackle this issue, we have developed guided-wave optical coherence elastography that can simultaneously excite and analyze symmetric (S0) and anti-symmetric (A0) elastic waves in the cornea at frequencies around 10 kHz and allows us to extract tensile and shear properties from measured wave dispersion curves. By applying acoustoelastic theory that incorporates corneal tension and a nonlinear constitutive tissue model, we verified the technique using elastomer phantoms and *ex vivo* porcine corneas and investigated the dependence on intraocular pressure. For two healthy human subjects, we measured a mean tensile modulus of 3.6 MPa and a mean shear modulus of 76 kPa *in vivo* with estimated errors of < 4%. This technique shows promise for the quantitative biomechanical assessment of the cornea in a clinical setting.

1. Introduction

The mechanical properties of the cornea play an essential role in establishing biomechanical homeosis with intraocular pressure (IOP) and maintaining normal corneal shapes. Measuring corneal biomechanics is significant in various aspects of corneal health and disease management, including the development of novel diagnostic metrics for early detection of keratoconus, monitoring corneal crosslinking, and accurately predicting corneal shapes after refractive surgeries.

The stromal tissue of the cornea has a lamellar microstructure [1-3], which make is mechanically anisotropic and approximately transverse isotropic [4]. Therefore, both tensile and shear modulus information is necessary to accurately describe corneal biomechanics. However, current techniques have severe limitations in measuring tensile and shear moduli *in vivo*. Mechanical tools, such as stress-strain tests and torsional tests, are invasive and not easily configurable for *in vivo* measurements. Two commercial instruments, namely the Ocular Response Analyzer (Reichert) [5] and Corvis ST (Oculus) [6, 7], provide phenomenological, biomechanical indices through the inverse analysis of corneal deformation against air-puffs but do not explicitly generate elastic modulus information. Corneal indentation can, in principle, measure tensile modulus, but with compromised accuracy [8, 9]. Brillouin microscopy can measure longitudinal modulus with high resolution [10, 11], but this elastic property is not directly related to tensile and shear properties required to describe corneal deformation. Ultrasound elastography [12] and optical coherence elastography (OCE) [13-17] have been applied to the cornea *in vivo*. To date, these techniques have employed almost exclusively the antisymmetric (A0) elastic wave, whose velocity is predominantly governed by out-of-plane shear modulus [18]. In-plane tensile moduli of the cornea has been estimated from the velocity dispersion [18] or the displacement profile [19] of the A0 wave, but with relatively large fitting uncertainties.

In this work, we demonstrate a noninvasive elastography method to quantify both the tensile and shear moduli of the cornea using guided elastic waves. We employ a high-frequency OCE technique to simultaneously generate symmetric (S0) as well as antisymmetric A0 guided waves in the cornea and measure their propagation speeds. We adopt an acoustoelastic wave model that incorporates IOP-induced tension, tissue anisotropy, and nonlinearities. We then obtain tensile and shear moduli directly by fitting the model to the measured dispersion curves of the S0 and A0 waves. After testing this technique using phantoms and porcine corneas, we apply it to *in vivo* human corneas.

2. Theory of Guided Elastic Waves

*2.1 Mechanical model of the cornea*

The deformation of tissue in response to mechanical stress is governed by its tensile and shear properties. The stroma dominates the overall stiffness of the cornea, accounting for approximately 90% of its total thickness [20]. The stroma consists of 200-300 lamellar layers containing collagen fibrils aligned along the layers (Fig. 1A), which are typically orthogonally stacked between adjacent lamellas [21]. The collagen fibers bear the tensile stress induced by the IOP [22, 23]. The in-plane tensile stress $\sigma$ can be estimated using the Young-Laplace equation:

$$\sigma = \text{IOP} \times R/(2h), \tag{1}$$

where $R$ and $h$ denote the radius and thickness of the cornea, respectively (Fig. 1B).

Given the corneal microstructure, it is reasonable to adopt the Holzapfel-Gasser-Ogden (HGO) model [24], which has been widely used for mechanical modeling of arterial walls. The strain energy function of the HGO model can be written as:

$$W = \frac{\mu}{2}(I_1 - 3) + \frac{k_1}{k_2}\sum_{i=1}^{2}\{e^{k_2[\kappa(I_1-3)+(1-3\kappa)(I_{4i}-1)]^2} - 1\}, \tag{2}$$

where $\mu$, $k_1$, $k_2$ and $\kappa$ are constitutive parameters. $\mu$ represents the initial shear modulus, $k_1$ represents the anisotropic tensile response, and $k_2$ describes the nonlinear stiffening effect of collagen fibrils at large strain. We assume $\kappa = 0$ since the collagen fibrils are aligned [24]. The influence of corneal curvature on wave speed is negligible (see Supplementary Fig. S1). Therefore, we can use Cartesian coordinates $(x_1, x_2, x_3)$ as shown in Fig. 1C. The axes of the collagen fibrils are denoted by $M = (1,0,0)^\text{T}$ and $M' = (0,0,1)^\text{T}$. The invariants in Eq. (2), $I_1$, $I_{41}$ and $I_{42}$, can be defined by $M$, $M'$, and the deformation gradient $F$: $I_1 = \text{tr}(F^TF)$, $I_{41} = (FM) \cdot (FM)$, $I_{42} = (FM') \cdot (FM')$ [24]. The Cauchy stress tensor can be determined using $\sigma_{ij} = F_{iI}\,\partial W/\partial F_{jI} - p\delta_{ij}$, where $p$ is a Lagrange multiplier for material incompressibility, $\delta_{ij}$ is the Kronecker delta, and the Einstein summation convention is used for $I \in \{1,2,3\}$. For biaxial stress $\sigma$, which corresponds to in-plane corneal tension, the deformation strain tensor is $F = \text{diag}(\lambda, \lambda^{-2}, \lambda)$, where $\lambda$ is the stretch ratio along $x_1$ and $\det(F) = 1$ due to the incompressibility of tissue. $\lambda$ is obtained from

$$\sigma = F_{1I}\frac{\partial W}{\partial F_{1I}} - F_{2I}\frac{\partial W}{\partial F_{2I}}. \tag{3}$$

To estimate corneal tension, we consider a typical corneal radius of curvature ($R$) of 7.5 mm and a corneal thickness ($h$) of 0.53 mm. For illustration, let us consider previously reported values of $\mu = 60$ kPa, $k_1 = 50$ kPa, $k_2 = 200$, and $\kappa = 0$ [25], although we find that this $k_1$ value is underestimated. Figure 1D illustrates the variations of stress $\sigma$ and stretch ratio $\lambda$ for a range of IOP from 0 to 40 mmHg. Due to the nonlinear mechanical properties, the stiffness of the cornea increases with tensile strain, and the stretch ratio $\lambda$ exhibits a decreasing slope as IOP increases.

We focus on the elastic waves guided in the cornea. In our OCE experiments, the vibration amplitudes of the elastic waves are only around 100 nm. Thus, the wave motion can be considered an incremental, linear perturbation on the prestressed configuration of the cornea, governed by the wave equation:

$$\alpha\frac{\partial^4\psi}{\partial x_1^4} + 2\beta\frac{\partial^4\psi}{\partial x_1^2 \partial x_2^2} + \gamma\frac{\partial^4\psi}{\partial x_2^4} = \rho\left(\frac{\partial^4\psi}{\partial x_1^2 \partial t^2} + \frac{\partial^4\psi}{\partial x_2^2 \partial t^2}\right), \tag{4}$$

where $\rho$ is the density, $\psi$ is related to the displacement components induced by wave motion ($u_1$ and $u_2$) as $u_1 = \partial\psi/\partial x_2$ and $u_2 = -\partial\psi/\partial x_1$. The coefficients $\alpha$, $\beta$, and $\gamma$ are determined by the strain energy function and $\lambda$ (see Refs. [26, 27] and Supplementary Note 1). The upper boundary (epithelium) interfacing with the air is stress-free, while the lower boundary (endothelium) is in contact with the aqueous humor. By solving Eq. (4) together with the linear acoustic equation of fluid (aqueous humor) and considering the appropriate boundary conditions, we obtain a secular equation that describes the wave velocity dispersion of the guided waves in the cornea (Supplementary Note 1).

*2.2 The guided waves and elastic moduli*

According to the Lamb wave theory [28, 29], the plate geometry of the cornea supports two fundamental guided waves with symmetric (S0) and antisymmetric (A0) motions, respectively. The S0 mode is an extensional, dilatational wave with its velocity largely governed by tensile modulus along the propagation direction. The A0 mode is a flexural, bending wave with its velocity governed by out-of-plane shear modulus.

Figure 1E shows the theoretical dispersion curves of the A0 and S0 waves as a function of frequency for two different IOP levels, 0 and 15 mmHg. At zero frequency, the velocity of the A0 wave is zero in the presence of the aqueous humor (nonzero without the fluid). The S0 wave speed in the low frequency limit is equal to $\sqrt{E^*/\rho}$, where $E^* = 4\mu + 4k_1$ is the in-plane tensile modulus (see Supplementary Note 2). Since $E^*$ is derived from the S0 wave, it corresponds to plane-strain tensile modulus. For highly anisotropic materials, such as the corneal tissue, $E^*$ is nearly equivalent to uniaxial tensile modulus (or Young's) modulus. On the other hand, in the high frequency regime where their wavelengths become shorter than the corneal thickness, both the A0 and S0 velocities converge to plateaus. At zero tension, the asymptotic velocities are close to $\sqrt{\mu/\rho}$ (see Supplementary Note 2). Corneal tension alters the high-frequency limit, which depends on both $\mu$ and $\sigma$ (see Supplementary Note 2). We introduce a ratio, $E^*/4\mu = 1 + k_1/\mu$, as an anisotropy index.

For illustration, in Fig. 1F we plot the phase velocities of the A0 mode at 10 kHz and the S0 mode at 4 kHz as a function of the IOP level for the modestly anisotropic material parameters ($\mu = 60$ kPa, $k_1 = 50$ kPa, $k_2 = 200$). We find that the phase velocity of A0 mode varies approximately linearly with IOP with a slope of ~ 0.04 m/s·mmHg ($r^2 > 0.99$). On the other hand, the S0 wave speed is nearly insensitive to IOPs lower than 10 mmHg, but increases with IOPs greater than 15 mmHg with a slope of ~ 0.91 m/s·mmHg ($r^2 > 0.99$). The S0 wave is more sensitive than the A0 wave to IOP in the physiological range because of the large exponential nonlinearity of tensile elasticity against the corneal tension. It should be noted that the tension affects the wave velocities not only through the nonlinear material properties but also directly through the tension-induced restoring forces on the wave displacements. The acoustoelastic analysis using $\alpha$, $\beta$, and $\gamma$ effectively decomposes the two mechanisms and allows us to extract the elastic moduli from the wave velocities (Supplementary Note 1).

## 3. Methods

*3.1 OCE system*

We used a custom-built, swept-source phase-sensitive optical coherence tomography (OCT) imaging system [13]. The OCT system has a center wavelength of 1300 nm and a bandwidth of 80 nm at an A-line rate of 43.2 kHz, offering an axial resolution of ~ 15 µm. The illumination power on the cornea is below 10 mW in compliance with the ANSI Z136.1-2014 safety standard. The displacements of the guided waves were measured using the method previously described [13, 29]. Briefly, the stimulus frequency was varied typically from 2 to 16 kHz with an interval of 2 kHz. The data acquisition time was about 0.4 s for each frequency. At each transverse location of the optical beam, 172 A-lines are acquired, which constitutes a single M-scan data. A total of 96 transverse positions along the sample are scanned. We extracted displacement profiles over time $t$ at each transverse location, and then performed a 1-dimensional Fourier transform to move the data from time $t$ domain to frequency $f$ domain. The frequency domain data was filtered at the driving frequency to reduce noise. After we obtained displacement profiles over the $x$ coordinate, 1-dimensional Fourier transform was applied to move the data from the spatial $x$ domain to the wavenumber $k_x$ domain to measure the wavenumber $k$ of each wave. This filtering in the $k_x$ domain is critical to remove other higher-order modes especially at high frequencies [30]. The phase velocity is then determined by $v = 2\pi f/k$.

*3.2 Optimization of the wave-excitation probe*

To generate guided waves in the cornea, we utilized a vibrating probe consisting of a piezoelectric transducer (PZT) and a probe tip [19], as depicted in Fig. 2B. The efficiency of wave excitation is maximized when the stress profile induced by the probe matches the stress profile of the targeted wave. Specifically, the contact length of the tip should be approximately half the wavelength of the wave of interest [31]. In order to optimize simultaneous excitation of both the A0 and S0 waves in the human cornea, we designed a flat tip with a contact length ($d$) of approximately 1.5 mm. We conducted experiments with different tilt angles ($\alpha$) for the probe, specifically 0, 15, and 30 degrees. Among these angles, we found that the A0 mode exhibited the highest vibration amplitude when the tip was tilted at an angle of 0 degrees ($\alpha = 0$ deg). On the other hand, the S0 mode was most efficiently excited with a tilt angle of 15 degrees ($\alpha = 15$ deg) (Supplementary Fig. S3). Considering that S0 wave dispersion data in mid to high frequencies are critical, we selected the tilt angle of 15 degrees and employed this configuration for all subsequent experiments.

## 4. Results

*4.1 Validation using phantoms*

To validate the OCE method, we conducted experiments using a 0.45-mm thick polydimethylsiloxane (PDMS) sheet, which is an isotropic material. The PDMS sheet was mounted in an artificial anterior chamber. Initially, no pressure or in-plane stress was applied to the PDMS sheet. Displacement maps and surface displacement profiles were obtained at 4 kHz and 12 kHz, as shown in Figs. 2C and 2D, respectively. In the wavenumber domain (Fig. 2E), two distinct peaks were observed. The peak at the lower wavenumber corresponded to the S0 mode with a higher phase velocity, while the peak at the higher wavenumber corresponded to the A0 mode with a lower velocity. The A0 wave was clearly detected across all frequencies ranging from 2 to 16 kHz. However, the S0 wave was reliably identified only at frequencies of 6 kHz and above but was not consistently observed at 2 and 4 kHz due to the relatively short contact length (1.5 mm) of the probe tip compared to the wavelengths at the low frequencies. The corresponding phase velocity dispersion curves are depicted in Fig. 2F, showing good agreement between the experimental data and our theoretical model. For this tension-free, isotropic PDMS phantom, we determined $\mu = 184$ kPa and $E^* = 736$ kPa. These values satisfy the expected relationship $E^* = 4\mu$, which holds only for isotropic materials.

To investigate the effect of intraocular pressure (IOP), we increased the IOP from 0 to 40 mmHg by attaching a water column with controlled height to the artificial anterior chamber. Figure 2G presents the A0 wave velocity at 16 kHz as a linear function of IOP, demonstrating a linear slope of 0.02 m/s·mmHg, or 0.2% increase per mmHg. This slope aligns with the theoretical expectation for a neo-Hookean model of PDMS ($\mu = 184$ kPa, $k_1 = 0$).

*4.2 Porcine corneas ex vivo*

In order to further validate the OCE method, we conducted experiments using fresh porcine eye globes *ex vivo*, where the IOP was controlled using a saline water column, as shown in Fig. 3A. Displacement maps were obtained at frequencies of 2 kHz, 4 kHz, and 6 kHz at an IOP of 5 mmHg, as depicted in Fig. 3B. Corresponding displacement profiles at the cornea surface are presented in Fig. 3C. In the wavenumber domain, both the A0 and S0 modes were clearly identified, with the S0 mode becoming dominant at higher frequencies. By fitting the measured dispersion curves over frequency (Fig. 3E), we determined an out-of-plane shear modulus of $\mu = 9.0 \pm 0.6$ kPa and an in-plane tensile modulus of $E_1^* = 216 \pm 22$ kPa. The ratio $E_1^*/4\mu$ was found to be 6, indicating a significant tensile-to-shear anisotropy in the corneal tissue. Figure 3F displays the wave velocities measured at 2 kHz and 10 kHz from seven porcine eye globe samples. The mean phase velocities for the A0 and S0 modes at 2 kHz were $2.18 \pm 0.08$ m/s and $17.9 \pm 6.1$ m/s (Mean ± SD), respectively, while the A0 wave velocity at 10 kHz was $3.46 \pm 0.19$ m/s.

The dependence of phase velocity on IOP was investigated by increasing the IOP from 5 mmHg up to 40 mmHg in increments of 5 or 10 mmHg. The deformation of the cornea, as measured from the OCT images, exhibited good agreement with our numerical simulations (Supplementary Fig. S4). Figure 3G illustrates the IOP-dependence of the A0 wave at 10 kHz, demonstrating a linear relationship. The velocity slope was found to be 0.12 m/s·mmHg or a 2.5% increase per mmHg at 15 mmHg. Figure 3H displays the IOP-dependence of the S0 wave at 2 kHz. The S0 wave velocity exhibited nonlinearity initially up to 10 mmHg, followed by a linear increase with larger IOPs. The slope at 15 mmHg was approximately 0.93 m/s·mmHg, or 3.3% increase per mmHg. The experimental data exhibited remarkable agreement with numerical simulation results based on the model with $\mu = 9$ kPa and $k_1 = 45$ kPa. A slightly higher value of 500 was used for $k_2$ to account for the nonlinearity of corneal tissues.

*4.3 Human corneas in vivo*

In the final phase of our study, we applied the OCE method to human eyes. Two healthy subjects were recruited: Subject 1 (31 years old male) and Subject 2 (62 years old male). The study was conducted at the Massachusetts General Hospital (MGH) with approval from the Institutional Review Board (IRB) of MGH and the Mass General Brigham Human Research Office. The excitation probe used was spring-loaded to maintain a small, constant force of <20 mN when in contact with the corneal surface (Fig. 4A). Prior to probe contact, a topical anesthetic was administered to the eye. Only the left eye of each subject was measured. A complete scan from 6 to 16 kHz, with a 2 kHz interval, took 2.4 seconds.

Figure 4B presents representative displacement profiles obtained from Subject 2. At 6 kHz, only the A0 mode was excited, while at 12 kHz and above, a combination of the S0 and A0 mode was observed. Measured surface wave displacement profiles are shown in Fig. 4C, and the corresponding wavenumber domain plots are displayed in Fig. 4D. At 16 kHz, peaks appearing in the negative wavenumber domain are likely caused by surface wave reflection presumably from the limbus and were therefore disregarded.

Figures 4E and 4F depict the measured phase velocity dispersion curves for the two human subjects. Each data point and error bar represent the mean and standard deviation of three consecutive OCE scans performed at the same location. The dispersion curves were fitted with the acoustoelastic model described in Section 2, assuming an IOP of 12 mmHg for both subjects. For Subject 1, we obtained an out-of-plane shear modulus $\mu = 81 \pm 3$ kPa and an in-plane tensile modulus $E_1^* = 3{,}728 \pm 57$ kPa. The ratio $E_1^*/4\mu$ was found to be 11.5. For subject 2, we determined an out-of-plane shear modulus $\mu = 70 \pm 3$ kPa and an in-plane tensile modulus $E_1^* = 3{,}400 \pm 138$ kPa, resulting in a ratio $E_1^*/4\mu = 12$. Our result revealed the significant mechanical anisotropy of *in vivo* human corneas, with an anisotropy ratio approximately twice that of the *ex vivo* porcine corneas at 10-15 mmHg.

## 5. Discussion

In this study, we presented a guided-wave OCE system that enabled simultaneous excitation and detection of the S0 and A0 waves in corneas. While the relationship between the waves and elastic moduli has been known, to our knowledge, this is the first experimental study that measures tensile modulus directly from the S0 wave in the cornea. The measured wave dispersion and dependence on pre-stress (corneal tension) showed good agreement with our acoustoelastic theory based on the HGO constitutive model. By fitting the phase velocity dispersion curves of the S0 and A0 waves using the model, we obtained the in-plane tensile modulus and out-of-plane shear modulus of human corneas. In terms of the model parameters, we measured $\mu$ = 70 - 80 kPa and $k_1$ = 770 - 880 kPa from two healthy subjects ($k_2 = 500$). The measured tensile modulus $E^*$ = 3.4 - 3.7 MPa was 47 times larger than the shear modulus $\mu$ for the two subjects with normal IOP.

Comparing our results with previous mechanical measurements on *ex vivo* corneal tissues, tensile moduli have been reported in the range of 0.2 to 3 MPa [32-34], while torsional tests applying shear stress along the corneal plane exhibited elastic moduli ranging from 3 to 50 kPa [35, 36]. Our *in vivo* results measured at ~ 10 kHz are slightly higher values compared to these mechanically measured values. We also compared our data with previous *in vivo* measurements reported in the literature (see Appendix Table A1). Some studies only reported group velocities [2, 15, 37, 38], which cannot quantity the elastic modulus since it depends on both the mechanical properties and the geometry of the cornea, such as thickness, due to waveguide dispersion. Our A0 wave speed values were reasonably consistent with the reported values from previous OCE studies [13, 16]. Our measured tensile moduli were significantly larger than previously reported values of <0.8 MPa using indentation and tonometer-based methods [8, 9], suggesting that the movement of the eyeball and variation of the IOP induced by indentation may have led to an underestimation of the tensile modulus.

Previous extensometry studies of cadaver corneal tissues have shown an age-dependent increase in tensile modulus [39], while our previous *in vivo* OCE work revealed an age-dependent decrease of shear modulus [13]. One plausible explanation for the opposite trend is that collagen fibers in the cornea lose elasticity and become stiffer with age, resulting in increased tensile modulus, while the interfibrillar matrix softens and diminishes shear modulus. In our current pilot study, the 62-year-old subject had 10% lower shear and tensile moduli compared to the 31-year-old subject. However, due to the limited number of samples, we cannot draw a statistically meaningful conclusion about age dependence. Follow-up clinical studies involving a large number of subjects across the lifespan are warranted to investigate the age dependence of the elastic properties.

To capture the full dispersion curve of the S0 wave in human corneas, it is necessary to extend the frequency range beyond 16 kHz. Additionally, patients with ocular hypertension are expected to have even higher S0 wave speeds, necessitating higher frequencies for accurately characterizing the S0 waves. We have recently developed an ultrahigh-frequency OCE method [40] and plan to incorporate this technique in our future clinical studies. With further optimizations, the guided-wave OCE technique holds promise for comprehensive studies of corneal mechanics and its role in the management of corneal diseases and refractive surgeries.

**Funding.** National Institutes of Health (NIH) via grants (R01-EY033356, P01-EY034857).

**Disclosures.** The authors declare that there are no conflicts of interest related to this article.

**Supplemental document.** See **Supplement 1** for supporting content.

# Appendix

**Table A1.** Previous *in vivo* measurements of corneal biomechanics

| Refs | Stimulus (Frequency) | Method | Subject | Results | Model |
|---|---|---|---|---|---|
| Sit [12] | Shaker (100 Hz) | US | Human (N = 20) | Phase velocity, 1.82 ± 0.10 m/s | Lamb wave |
| Nguyen [2] | ARF (~800 Hz) | US | Porcine (N = 4) | Group velocity, 6 ~ 8 m/s | Lamb wave |
| Nguyen [37] | ARF (~800 Hz) | US | Porcine (N = 4) | Group velocity, ~1.9 m/s | Lamb wave |
| Ramier [13] | Contact probe (up to 16 kHz) | OCE | Human (N = 12) | Phase velocity, 7.86 ± 0.75 m/s | Rayleigh surface wave |
| Lan [15] | Air-puff (~500 Hz) | OCE | Human (N = 18) | Group velocity, ~3.5 m/s | - |
| Jin [16] | Air-puff (~500 Hz) | OCE | Human (N = 12) | Phase velocity, 12.73 ± 1.46 m/s (>2 kHz) | Scholte wave |
| Zvietcovich [41] | ARF (2 kHz) | OCE | Rabbit (N = 4) | Phase velocity, 3.75 ~ 5.72 m/s | Lamb wave |
| Mekonnen [42] | ARF (<3 kHz) | OCE | Rabbit (N = 4) | Phase velocity, 3.41 ± 0.52 m/s | Lamb wave |
| Kirby [18] | ARF (<3 kHz) | OCE | Rabbit (N = 4) | Shear modulus 34 kPa to 261 kPa and tensile modulus 20 MPa to 44 MPa | Lamb wave |
| Jin [43] | ARF (~500 Hz) | OCE | Rabbit (N = 5) | Phase velocity, 7.52 ± 1.11 m/s | Rayleigh surface wave |
| Li [38] | ARF (~500 Hz) | OCE | Rabbit (N = 4) | Group velocity, ~6.2 m/s | - |
| Kling [6] | Air-puff (~30 Hz) | Corvis ST | Human (N = 9) | Young's modulus, ~0.40 MPa for posterior and ~1.52 MPa for anterior cornea | Finite element analysis |

| Shih [7] | Air-puff (~30 Hz) | Corvis ST | Human (N = 25) | Young's modulus, ~0.32 MPa | Modal analysis |
| Pye [8] | Applanation (Static) | Applanation Tonometry | Human (N = 100) | Young's modulus, 0.25 ± 0.1 MPa | Orssengo-Pye model |
| Lam [9] | Flat indenter (Static) | Indentation | Human (N = 29) | Tangent modulus, 0.755 ± 0.159 MPa | Thin shell model |

ARF: acoustic radiation force, US: ultrasound elastography, OCE: optical coherence elastography

**References**

[1] A. Elsheikh, D. Alhasso, Mechanical anisotropy of porcine cornea and correlation with stromal microstructure, Experimental eye research 88(6) (2009) 1084-1091.

[2] T.-M. Nguyen, J.-F. Aubry, M. Fink, J. Bercoff, M. Tanter, In vivo evidence of porcine cornea anisotropy using supersonic shear wave imaging, Investigative ophthalmology & visual science 55(11) (2014) 7545-7552.

[3] A.M. Eltony, Peng Shao, and Seok-Hyun Yun, Measuring mechanical anisotropy of the cornea with Brillouin microscopy, Nature communications 13 (2022) 1354.

[4] J.J. Pitre, M.A. Kirby, D.S. Li, T.T. Shen, R.K. Wang, M. O'Donnell, I. Pelivanov, Nearly-incompressible transverse isotropy (NITI) of cornea elasticity: model and experiments with acoustic micro-tapping OCE, Scientific reports 10 (2020) 12983.

[5] D.A. Luce, Determining in vivo biomechanical properties of the cornea with an ocular response analyzer, Journal of Cataract & Refractive Surgery 31(1) (2005) 156-162.

[6] S. Kling, N. Bekesi, C. Dorronsoro, D. Pascual, S. Marcos, Corneal viscoelastic properties from finite-element analysis of in vivo air-puff deformation, PloS one 9(8) (2014) e104904.

[7] P.J. Shih, H.J. Cao, C.J. Huang, I.J. Wang, W.P. Shih, J.Y. Yen, A corneal elastic dynamic model derived from Scheimpflug imaging technology, Ophthalmic and Physiological Optics 35(6) (2015) 663-672.

[8] D.C. Pye, A clinical method for estimating the modulus of elasticity of the human cornea in vivo, Plos one 15(1) (2020) e0224824.

[9] A.K. Lam, Y. Hon, L.K. Leung, D.C. Lam, Repeatability of a novel corneal indentation device for corneal biomechanical measurement, Ophthalmic and Physiological Optics 35(4) (2015) 455-461.

[10] P. Shao, A.M. Eltony, T.G. Seiler, B. Tavakol, R. Pineda, T. Koller, T. Seiler, S.-H. Yun, Spatially-resolved Brillouin spectroscopy reveals biomechanical abnormalities in mild to advanced keratoconus in vivo, Scientific reports 9 (2019) 7467.

[11] H. Zhang, L. Asroui, I. Tarib, W.J. Dupps, G. Scarcelli, J.B. Randleman, Motion Tracking Brillouin Microscopy Evaluation of Normal, Keratoconic, and Post-Laser Vision Correction Corneas: Motion Tracking Brillouin Microscopy in Keratoconus and Laser Vision Correction, American Journal of Ophthalmology 254 (2023) 128-140.

[12] A.J. Sit, S.-C. Lin, A. Kazemi, J.W. McLaren, C.M. Pruet, X. Zhang, In Vivo Non-Invasive Measurement of Young's Modulus of Elasticity in Human Eyes: A Feasibility Study, Journal of glaucoma 26(11) (2017) 967.

[13] A. Ramier, A.M. Eltony, Y. Chen, F. Clouser, J.S. Birkenfeld, A. Watts, S.-H. Yun, In vivo measurement of shear modulus of the human cornea using optical coherence elastography, Scientific reports 10 (2020) 17366.

[14] F. Zvietcovich, J. Birkenfeld, A. Varea, A.M. Gonzalez, A. Curatolo, S. Marcos, Multi-meridian wave-based corneal optical coherence elastography in normal and keratoconic patients, Investigative Ophthalmology & Visual Science, 2022, pp. 2380–A0183.

[15] G. Lan, S.R. Aglyamov, K.V. Larin, M.D. Twa, In vivo human corneal shear-wave optical coherence elastography, Optometry and Vision Science 98(1) (2021) 58.

[16] Z. Jin, S. Chen, Y. Dai, C. Bao, S. Ye, Y. Zhou, Y. Wang, S. Huang, Y. Wang, M. Shen, In vivo noninvasive measurement of spatially resolved corneal elasticity in human eyes using Lamb wave optical coherence elastography, Journal of Biophotonics 13(8) (2020) e202000104.

[17] Y. Li, J. Zhu, J.J. Chen, J. Yu, Z. Jin, Y. Miao, A.W. Browne, Q. Zhou, Z. Chen, Simultaneously imaging and quantifying in vivo mechanical properties of crystalline lens and cornea using optical coherence elastography with acoustic radiation force excitation, APL photonics 4(10) (2019) 106104.

[18] M.A. Kirby, I. Pelivanov, G. Regnault, J.J. Pitre, R.T. Wallace, M. O'Donnell, R.K. Wang, T.T. Shen, Acoustic micro-tapping optical coherence elastography to quantify corneal collagen cross-linking: an ex vivo human study, Ophthalmology Science 3(2) (2023) 100257.

[19] G.-Y. Li, X. Feng, S.-H. Yun, In vivo optical coherence elastography reveals spatial variation and anisotropy of corneal stiffness, arXiv preprint arXiv:2307.04083 (2023).

[20] D.W. DelMonte, T. Kim, Anatomy and physiology of the cornea, Journal of Cataract & Refractive Surgery 37(3) (2011) 588-598.

[21] K.M. Meek, A.J. Quantock, The use of X-ray scattering techniques to determine corneal ultrastructure, Progress in retinal and eye research 20(1) (2001) 95-137.

[22] J.W. Ruberti, A. Sinha Roy, C.J. Roberts, Corneal biomechanics and biomaterials, Annual review of biomedical engineering 13 (2011) 269-295.

[23] A. Kotecha, What biomechanical properties of the cornea are relevant for the clinician?, Survey of ophthalmology 52(6) (2007) S109-S114.

[24] T.C. Gasser, R.W. Ogden, G.A. Holzapfel, Hyperelastic modelling of arterial layers with distributed collagen fibre orientations, Journal of the royal society interface 3(6) (2006) 15-35.

[25] A. Pandolfi, Cornea modelling, Eye and Vision 7 (2020) 2.

[26] M. Destrade, G. Saccomandi, Waves in nonlinear pre-stressed materials, Springer Science & Business Media 2007.

[27] G.-Y. Li, Q. He, R. Mangan, G. Xu, C. Mo, J. Luo, M. Destrade, Y. Cao, Guided waves in pre-stressed hyperelastic plates and tubes: Application to the ultrasound elastography of thin-walled soft materials, Journal of the Mechanics and Physics of Solids 102 (2017) 67-79.

[28] Z.L. Han, J.S. Li, M. Singh, C. Wu, C.H. Liu, R. Raghunathan, S.R. Aglyamov, S. Vantipalli, M.D. Twa, K.V. Larin, Optical coherence elastography assessment of corneal viscoelasticity with a modified Rayleigh-Lamb wave model, J. Mech. Behav. Biomed. Mater. 66 (2017) 87-94.

[29] A. Ramier, B. Tavakol, S.-H. Yun, Measuring mechanical wave speed, dispersion, and viscoelastic modulus of the cornea using optical coherence elastography, Opt Express 27(12) (2019) 16635-16649.

[30] G.-Y. Li, X. Feng, A. Ramier, S.-H. Yun, Supershear surface waves reveal prestress and anisotropy of soft materials, Journal of the Mechanics and Physics of Solids 169 (2022) 105085.

[31] X. Feng, G.-Y. Li, A. Ramier, A.M. Eltony, S.-H. Yun, In vivo stiffness measurement of epidermis, dermis, and hypodermis using broadband Rayleigh-wave optical coherence elastography, Acta Biomaterialia 146 (2022) 295-305.

[32] A. Elsheikh, D. Wang, M. Brown, P. Rama, M. Campanelli, D. Pye, Assessment of corneal biomechanical properties and their variation with age, Current eye research 32(1) (2007) 11-19.

[33] G. Wollensak, E. Spoerl, T. Seiler, Stress-strain measurements of human and porcine corneas after riboflavin–ultraviolet-A-induced cross-linking, Journal of Cataract & Refractive Surgery 29(9) (2003) 1780-1785.

[34] N.E.K. Cartwright, J.R. Tyrer, J. Marshall, Age-related differences in the elasticity of the human cornea, Investigative ophthalmology & visual science 52(7) (2011) 4324-4329.

[35] S.J. Petsche, D. Chernyak, J. Martiz, M.E. Levenston, P.M. Pinsky, Depth-dependent transverse shear properties of the human corneal stroma, Investigative ophthalmology & visual science 53(2) (2012) 873-880.

[36] H. Hatami-Marbini, Viscoelastic shear properties of the corneal stroma, Journal of biomechanics 47(3) (2014) 723-728.

[37] T.-M. Nguyen, J.-F. Aubry, D. Touboul, M. Fink, J.-L. Gennisson, J. Bercoff, M. Tanter, Monitoring of cornea elastic properties changes during UV-A/riboflavin-induced corneal collagen cross-linking using supersonic shear wave imaging: a pilot study, Investigative ophthalmology & visual science 53(9) (2012) 5948-5954.

[38] Y. Li, S. Moon, J.J. Chen, Z. Zhu, Z. Chen, Ultrahigh-sensitive optical coherence elastography, Light: Science & Applications 9 (2020) 58.

[39] W.M. Petroll, M. Miron-Mendoza, Mechanical interactions and crosstalk between corneal keratocytes and the extracellular matrix, Experimental eye research 133 (2015) 49-57.

[40] X. Feng, G.-Y. Li, S.-H. Yun, Ultra-wideband optical coherence elastography from acoustic to ultrasonic frequencies, arXiv preprint arXiv:2211.10534 (2022).

[41] F. Zvietcovich, A. Nair, M. Singh, S.R. Aglyamov, M.D. Twa, K.V. Larin, In vivo assessment of corneal biomechanics under a localized cross-linking treatment using confocal air-coupled optical coherence elastography, Biomedical Optics Express 13(5) (2022) 2644-2654.

[42] T. Mekonnen, X. Lin, C. Zevallos-Delgado, M. Singh, S.R. Aglyamov, V. Coulson-Thomas, K.V. Larin, Longitudinal assessment of the effect of alkali burns on corneal biomechanical properties using optical coherence elastography, Journal of Biophotonics 15(8) (2022) e202200022.

[43] Z. Jin, R. Khazaeinezhad, J. Zhu, J. Yu, Y. Qu, Y. He, Y. Li, T.E.G. Alvarez-Arenas, F. Lu, Z. Chen, In-vivo 3D corneal elasticity using air-coupled ultrasound optical coherence elastography, Biomedical Optics Express 10(12) (2019) 6272-6285.

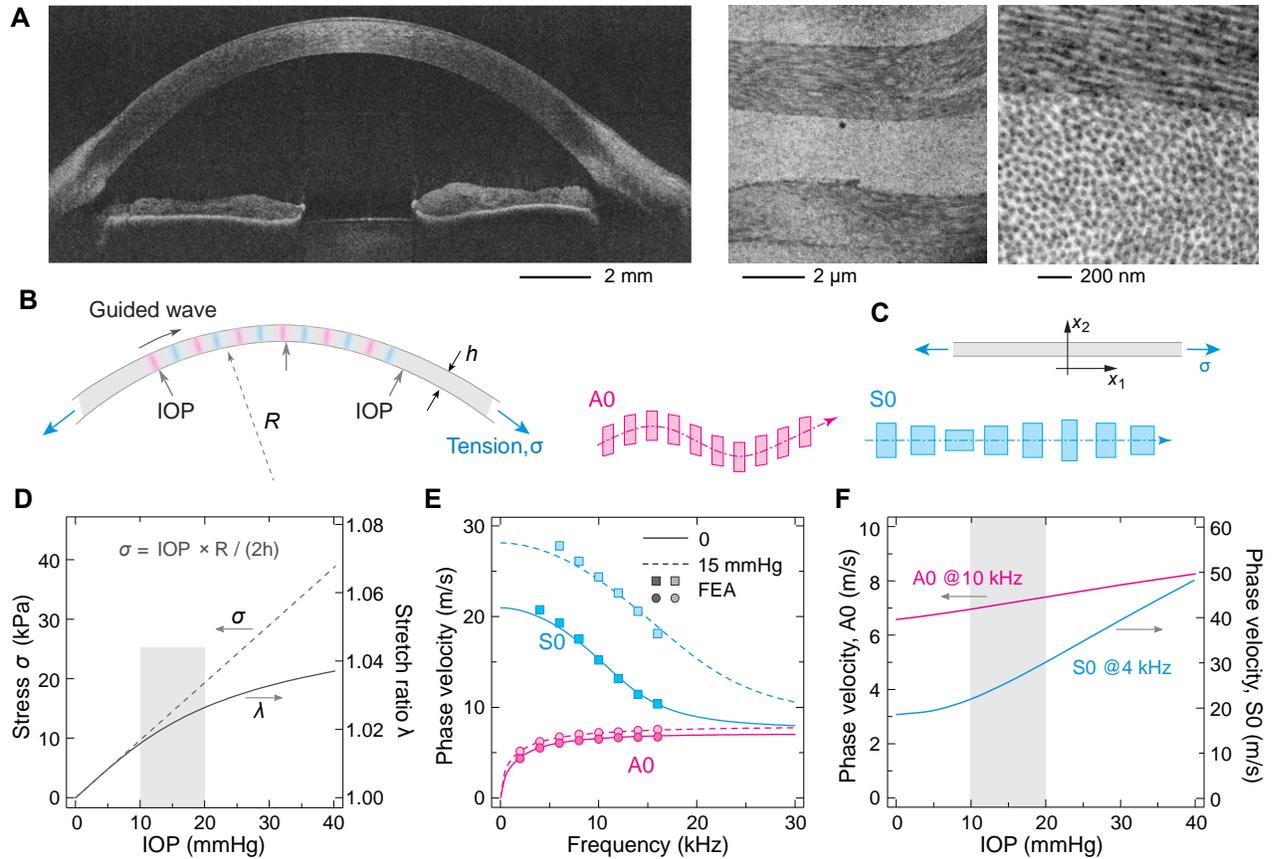

**Fig. 1. Mechanical model for elastic wave motion in the cornea.** (A) OCT image of a healthy volunteer (left), and electron micrographs of a porcine corneal tissue showing the microstructure of the stroma layer (center and right). (B) Static load of the cornea in physiological condition and the elastic wave motion. The fundamental guided modes are asymmetric (A0) and symmetric (S0), respectively. Boxes illustrate specific tissue deformations involved in the wave motion. (C) The coordinate system used in the model. (D) Stress $\sigma$ and stretch ratio $\lambda$ of the cornea under different IOP levels. The normal physiological range of the IOP is typically 10-20 mmHg (gray area). (E) Theoretical dispersion relations of the A0 and S0 modes at zero IOP (solid lines) and 15 mmHg (dashed lines). The markers show the results obtained from finite element analysis (FEA). (F) Effect of the IOP on the phase velocities of A0 (10 kHz) and S0 (4 kHz) modes for a weakly anisotropic material with $R = 7.5$ mm, $h = 0.53$ mm, $\mu = 30$ kPa, $k_1 = 50$ kPa, $k_2 = 200$, and $\kappa = 0$.

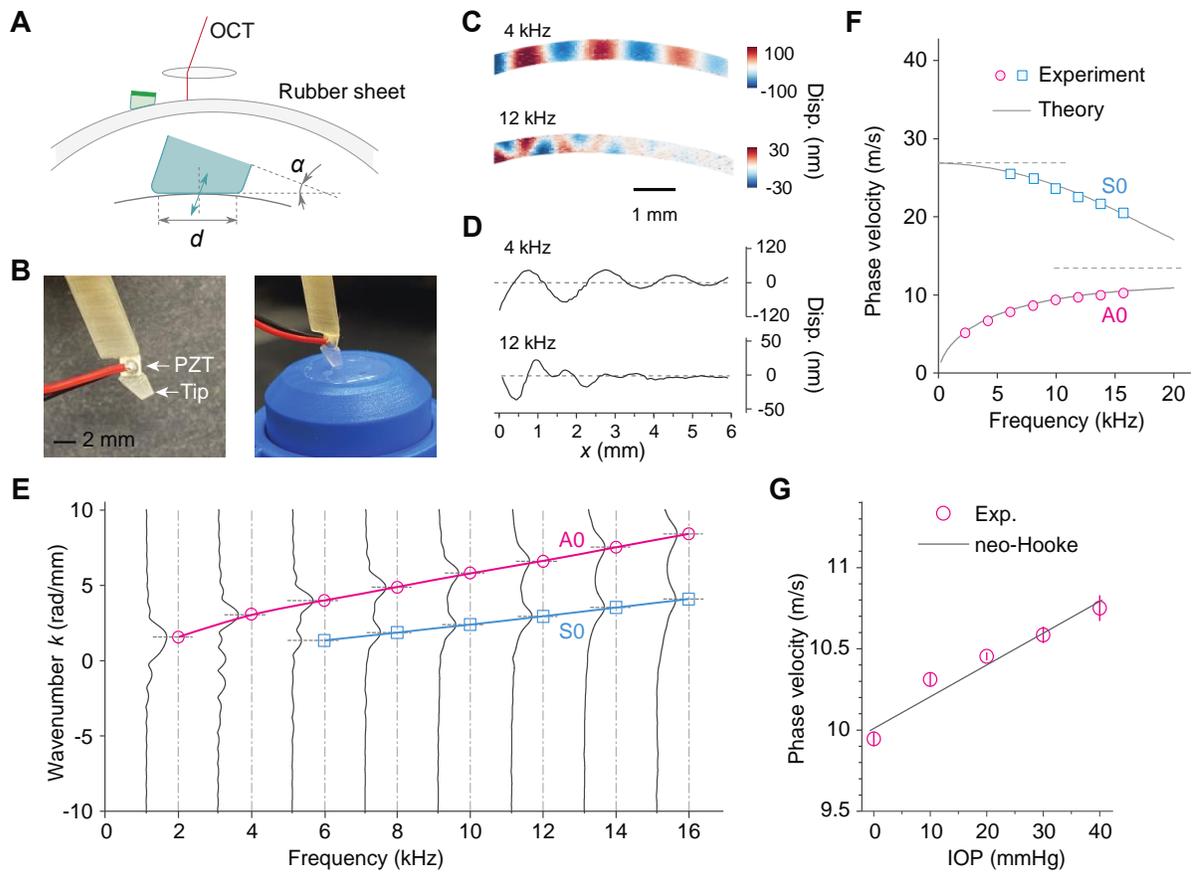

**Fig. 2. Verification using elastomer phantoms.** (A) Schematic of the experimental setup including the contact piezoelectric probe with a contact length $d$ and a tilt angle $\alpha$ between the vibration direction and the surface normal. (B) Pictures of the contact probe (left) and a PDMS sheet phantom on an artificial anterior chamber (right). (C) Displacement maps at 4 kHz and 12 kHz (IOP = 0 mmHg). (D) Corresponding surface displacement profiles. Only the real parts are shown. (E) Fourier transformations of the surface displacement profiles to measure the dispersion relations. The S0 mode is clearly identified at 6 -16 kHz. (F) Phase velocity dispersion relations and the theoretical model. (G) Effect of IOP on the A0 wave velocity at 16 kHz (n = 3 measurements on 3 different locations of one phantom). The line represents the theoretical fit based on the neo-Hookean model ($k_1 = 0$). Data are represented as mean values +/- SD.

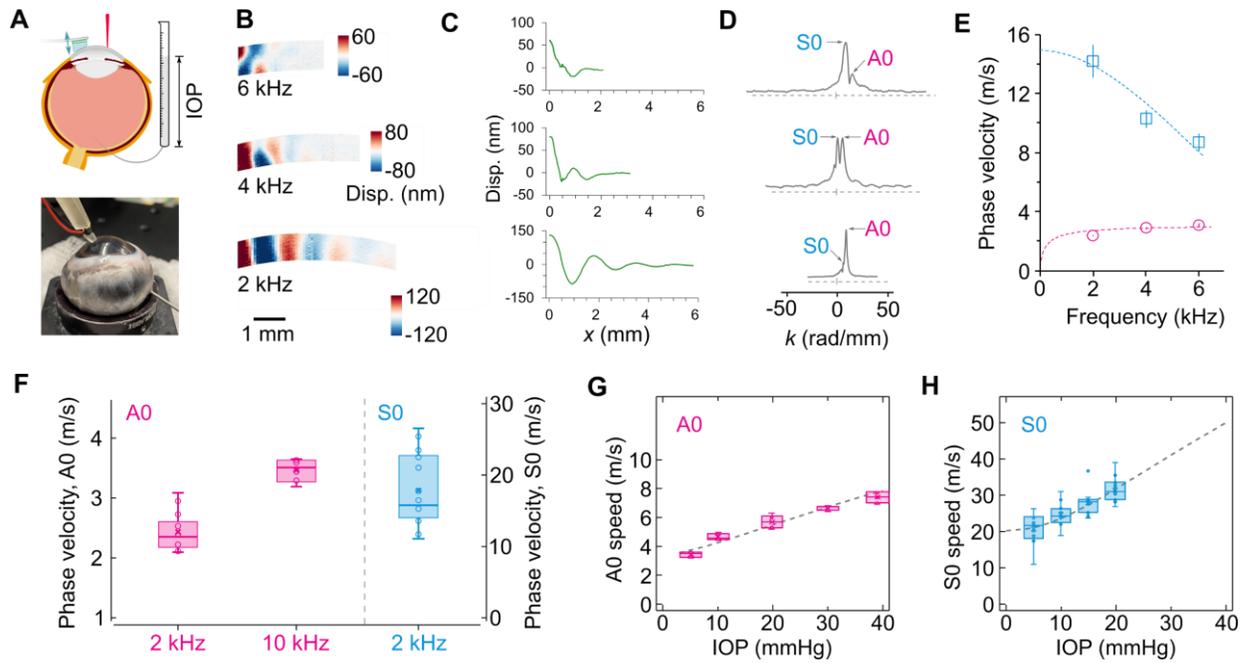

**Fig. 3. Simultaneous measurement of the A0 and S0 waves in *ex vivo* porcine corneas.** (A) Schematic of the experimental setup. IOP is controlled using a water column. (B) Maps of the displacement field. (C) Surface displacement extracted from the surface of the cornea at 2, 4, and 6 kHz (bottom to top). (D) Wavenumber domain profiles of the displacements and the identified A0 and S0 modes. (E) Corresponding dispersion relations of the A0 and S0 waves (n = 3 samples). Data are represented as mean values +/- SD. (F) Inter-sample variations of the A0 and S0 waves (n = 7 samples). (G) Effect of the IOP on the phase velocities of the A0 mode at 10 kHz (n = 7 samples). (H) Effect of the IOP on the phase velocities of the S0 mode at 2 kHz (n = 7 samples). Dashed curves represent the theoretical fit. For the boxplots In Fig. F, G, and H, the center line represents the median value, the lower and upper hinges correspond to the first and third quartiles (the 25th and 75th percentiles), the whisker extends from the hinge to the largest or smallest value at most 1.5 * interquartile range of the hinge.

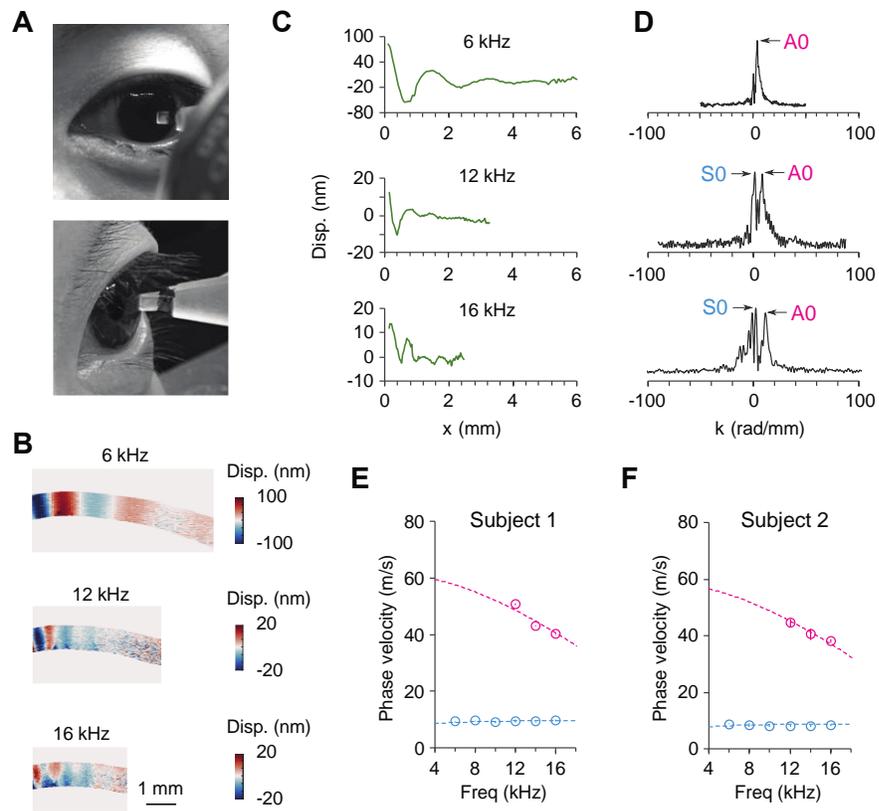

**Fig. 4. *In vivo* human measurement of the tensile and shear moduli using the S0 and A0 waves.** (A) Monitoring camera view of a human subject during measurement. (B) Displacement profiles at 6, 12, and 16 kHz for Subject 2. (C) Surface displacement. (D) Wavenumber domain Fourier transform result of the wave displacement. Arrows point to A0 and S0 modes. (E) Dispersion relations measured from Subject 1. (F) Dispersion relations measured from Subject 2. Dashed curves represent theoretical fits. For each subject, three measurements were taken at approximately the same location on the central cornea. Data are represented as mean values +/- SD.

# Supplementary materials

# Simultaneous tensile and shear measurement of the human cornea *in vivo* using S0- and A0-wave optical coherence elastography

Guo-Yang Li[1,†,#], Xu Feng[1,†], Seok-Hyun Yun[1,2,*]

[1] Harvard Medical School and Wellman Center for Photomedicine, Massachusetts General Hospital, 50 Blossom St., Boston, MA 02114, USA.

[2] Harvard-MIT Division of Health Sciences and Technology, Cambridge, MA 02139, USA.

[#] Current address: Department of Mechanical Engineering, Peking University, Peking, China

[†] Equal contribution

*Correspondence: syun@hms.harvard.edu

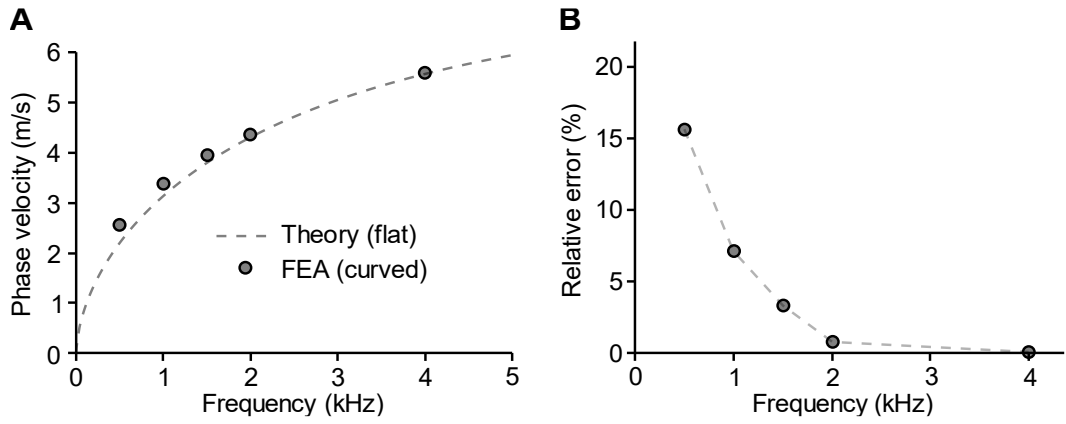

**Fig. S1** Influence of corneal curvature on wave speed. (A) Phase velocity dispersion relation. Dashed line: theoretical model assuming the cornea is flat. Circles: Finite element analysis (FEA) simulation considering corneal curvature. (B) Relative error of the phase velocities between the theory for the flat cornea and the FEA simulation of the curved cornea. The relative error becomes negligible at 4 kHz, as the wavelength is significantly smaller than the radius of curvature.

## Supplementary Note S1. Acoustoelastic model for guided waves in the cornea

The wave equation for small-amplitude plane elastic wave in a prestressed solid, which can be found in Ref. (*1*), is

$$\alpha \frac{\partial^4 \psi}{\partial x_1^4} + 2\beta \frac{\partial^4 \psi}{\partial x_1^2 \partial x_2^2} + \gamma \frac{\partial^4 \psi}{\partial x_2^4} = \rho \left( \frac{\partial^4 \psi}{\partial x_1^2 \partial t^2} + \frac{\partial^4 \psi}{\partial x_2^2 \partial t^2} \right), \tag{S1}$$

where the stream function $\psi$ is related to the displacement components $u_1$ and $u_2$ via the relation of $u_1 = \psi_{,2}$ and $u_2 = -\psi_{,1}$. These relations promise $u_{1,1} + u_{2,2} = 0$, which is equivalent to the material incompressible constraint adopted in this study. $\rho$ and $t$ denote the density and time, respectively. The coefficients $\alpha$, $\beta$, and $\gamma$ are determined by the constitutive law and the stretch ratio $\lambda$,

$$\alpha = \mathcal{A}^0_{1212}, 2\beta = \mathcal{A}^0_{1111} + \mathcal{A}^0_{2222} - 2\mathcal{A}^0_{1122} - 2\mathcal{A}^0_{1221}, \gamma = \mathcal{A}^0_{2121}, \tag{S2}$$

where the fourth-order tensor, $A_{0ijkl}$, is the Eulerian elasticity tensor and is defined as

$$\mathcal{A}^0_{ijkl} = F_{iI} F_{kJ} \frac{\partial^2 W}{\partial F_{jI} \partial F_{lJ}}, \quad i, j, k, l, I, J \in \{1,2,3\} \tag{S3}$$

where $\mathbf{F}$ is the deformation gradient tensor and $W$ is the strain energy function. In Eq. (S3) the Einstein summation convention is adopted. For the HGO model, the strain energy is

$$W = \frac{\mu}{2}(I_1 - 3) + \frac{k_1}{k_2} \sum_{i=1}^{2} \left\{ e^{k_2 [\kappa(I_1-3)+(1-3\kappa)(I_{4i}-1)]^2} - 1 \right\}, \tag{S4}$$

where $\mu$, $k_1$, $k_2$ and $\kappa$ are constitutive parameters. $\mu$ denotes the initial shear modulus. The dimension of $k_1$ is the same as $\mu$, whereas $k_2$ is a dimensionless parameter which determines the nonlinear hardening effect of the collagen fibrils when being stretched. $\kappa = 0$ if the collagen fibrils are ideally aligned (*2*), which is appliable for the cornea. $I_1 = \mathrm{tr}(\mathbf{F}^T \mathbf{F})$. $I_{41}$ and $I_{42}$ are two invariants related to two families of collagen fibers. Following the coordinate system $(x_1, x_2, x_3)$ shown in Fig. 1C, the axes of the collagen fibers of the cornea, denoted by unit vectors $\mathbf{M}$ and $\mathbf{M}'$, are aligned with $x_1$ and $x_3$, i.e., $\mathbf{M} = (1,0,0)^T$ and $\mathbf{M}' = (0,0,1)^T$. Then $I_{41}$ and $I_{42}$ can be determined by $\mathbf{M}$ and $\mathbf{M}'$ (*2*)

$$I_{41} = (\mathbf{FM}) \cdot (\mathbf{FM}), \quad I_{42} = (\mathbf{FM}') \cdot (\mathbf{FM}'). \tag{S5}$$

With the strain energy function and the deformation tensor, the Cauchy stress tensor can be determined by

$$\sigma_{ij} = F_{iI} \partial W / \partial F_{jI} - \bar{p} \delta_{ij}, \tag{S6}$$

where $\bar{p}$ is a Lagrange multiplier for material incompressibility and $\delta_{ij}$ is the Kronecker delta.

In this study, we consider a biaxial stretch, $\mathbf{F} = \mathrm{diag}(\lambda, \lambda^{-2}, \lambda)$, where $\lambda$ is the stretch ratio along $x_1$ (or $x_3$) axis. With $\sigma_{22} = 0$ we can obtain ($\sigma = \sigma_{11} = \sigma_{33}$)

$$\sigma = \mu(\lambda^2 - \lambda^{-4}) + 2k_1 \lambda^2 (\lambda^2 - 1) e^{\left[ k_2 (\lambda^2 - 1)^2 \right]}. \tag{S7}$$

According to the Young-Laplace equation, $\sigma = \mathrm{IOP} \times R / (2h)$. Therefore, the stretch ratio $\lambda$ can be obtained by solving the nonlinear equation

$$\mu(\lambda^2 - \lambda^{-4}) + 2k_1\lambda^2(\lambda^2 - 1)e^{[k_2(\lambda^2-1)^2]} = \text{IOP} \times R/(2h). \tag{S8}$$

The coefficients $\alpha$, $\beta$, and $\gamma$ can be obtained by inserting Eqs. (S3) and (S4) into Eq. (S2),

$$\alpha = \lambda^2\left\{\mu + 2k_1(\lambda^2 - 1)e^{[k_2(\lambda^2-1)^2]}\right\}, \quad \gamma = \mu\lambda^{-4}. \tag{S9}$$

$$2\beta = \alpha + \gamma + 4k_1\lambda^4[2k_2(\lambda^2 - 1)^2 + 1]e^{[k_2(\lambda^2-1)^2]},$$

In the absence of prestress (i.e., $\lambda = 1$)

$$\alpha = \mu, \beta = \mu + 2k_1, \gamma = \mu. \tag{S10}$$

Next, we consider the guided wave motion in the cornea. It should be noted the boundary conditions (BCs) have pronounced effects on the dispersion relations of the guided waves. The two sides of the cornea are air and aqueous humor. In our simplified model (see Fig. 1C), the aqueous humor is modeled as a semi-infinite fluid layer and the wave equation is

$$\nabla^2 \chi = \frac{\rho^f}{\nu}\chi_{,tt}, \tag{S11}$$

where $\nu$ (2.2 GPa) and $\rho^f$ (1,000 kg/m³) denote the bulk modulus and density of the fluid, respectively. The potential function $\chi$ is related to the displacement of the fluid (denoted by $\mathbf{u}^f$) via the relations of $u_1^f = \chi_{,1}$ and $u_2^f = \chi_{,2}$. The pressure of the fluid, denoted by $p^*$, is determined by

$$p^* = -\nu \nabla \cdot \mathbf{u}^f. \tag{S12}$$

At the interreference ($x_2 = 0$), the following interfacial conditions apply

$$u_2 = u_2^f, \quad -\gamma\psi_{,11} + \gamma\psi_{,22} = 0, \quad \rho\psi_{,2tt} - (2\beta + \gamma)\psi_{,112} - \gamma\psi_{,222} = -p_{,1}^*. \tag{S13}$$

At $x_2 = h$, the boundary is stress free, which gives

$$-\gamma\psi_{,11} + \gamma\psi_{,22} = 0, \quad \rho\psi_{,2tt} - (2\beta + \gamma)\psi_{,112} - \gamma\psi_{,222} = 0. \tag{S14}$$

More details on the derivations of the boundary conditions can be found in Ref. (*3, 4*).

We seek the plane wave solutions for $\psi(x_1, x_2, t)$ and $\chi(x_1, x_2, t)$, i.e.,

$$\begin{cases} \chi(x_1, x_2, t) = \chi_0(x_2)e^{\iota k(x_1 - ct)} \\ \psi(x_1, x_2, t) = \psi_0(x_2)e^{\iota k(x_1 - ct)} \end{cases} \tag{S15}$$

where $\iota = \sqrt{-1}$, $k$ is the wavenumber, and $c$ is the phase velocity. Inserting Eq. (S15) into Eqs. (S1) and (S11), we can get

$$\begin{cases} \chi = Ae^{-\xi k x_2}e^{\iota k(x_1 - ct)} \\ \psi = [B_1\cosh(s_1 k x_2) + B_2\sinh(s_1 k x_2) + B_3\cosh(s_2 k x_2) + B_4\sinh(s_2 k x_2)]e^{\iota k(x_1 - ct)} \end{cases} \tag{S16}$$

The parameters $s_1$, $s_2$ and $\xi$ are determined by

$$\gamma s^4 - (2\beta - \rho c^2)s^2 + (\alpha - \rho c^2) = 0, \tag{S17}$$

and

$$\xi^2 - 1 = -c^2 \rho^f / \nu. \tag{S18}$$

Substituting $\psi(x_1, x_2, t)$ and $\chi(x_1, x_2, t)$ in Eqs. (S13) and (S14) in terms of Eq. (S16) we get

$$M_{5 \times 5} \cdot [B_1, B_2, B_3, B_4, A]^{\mathrm{T}} = 0, \tag{S19}$$

where the nonzero components of $\mathbf{M}$ are

$$M_{11} = s_1^2 + 1, M_{13} = s_2^2 + 1,$$

$$M_{22} = \gamma s_1 (s_2^2 + 1), M_{24} = \gamma s_2 (s_1^2 + 1), M_{25} = \iota \rho^f c^2,$$

$$M_{31} = 1, M_{33} = 1, M_{35} = -\iota \xi,$$

$$M_{41} = (s_1^2 + 1)\cosh(s_1 k h), M_{42} = (s_1^2 + 1)\sinh(s_1 k h), \tag{S20}$$

$$M_{43} = (s_2^2 + 1)\cosh(s_2 k h), M_{44} = (s_2^2 + 1)\sinh(s_2 k h),$$

$$M_{51} = s_1(s_2^2 + 1)\sinh(s_1 k h), M_{52} = s_1(s_2^2 + 1)\cosh(s_1 k h),$$

$$M_{53} = s_2(s_1^2 + 1)\sinh(s_2 k h), M_{54} = s_2(s_1^2 + 1)\cosh(s_1 k h).$$

In this derivation, we use the identity

$$2\beta - \rho c^2 = \gamma(s_1^2 + s_2^2), \tag{S21}$$

which can be obtained from Eq. (S17). The existence of nontrivial solution to Eq. (S12) requires

$$\det(\mathbf{M}_{5 \times 5}) = 0, \tag{S22}$$

which is the secular equation for the guided waves in the cornea.

**Reference**

1. M. Destrade and G. Saccomandi, Waves in nonlinear pre-stressed materials. *Springer Science & Business Media*. **495**, 1–26 (2007).

2. T. C. Gasser, R. W. Ogden, G. A. Holzapfel, Hyperelastic modelling of arterial layers with distributed collagen fibre orientations. *J. R. Soc. Interface*. **3**, 15–35 (2005).

3. G.-Y. Li, Q. He, R. Mangan, G. Xu, C. Mo, J. Luo, M. Destrade, Y. Cao, Guided waves in pre-stressed hyperelastic plates and tubes: Application to the ultrasound elastography of thin-walled soft materials. *J. Mech. Phys. Solids*. **102** (2017).

4. M. Otténio, M. Destrade, R. W. Ogden, Acoustic waves at the interface of a pre-stressed incompressible elastic solid and a viscous fluid. *Int. J. Non. Linear. Mech.* **42**, 310–320 (2007).

## Supplementary Note 2. Phase velocities of A₀ and S₀ modes at high/low frequency regimes

When the wave frequency approaches to the infinity, the dispersion relations of A0 and S0 exhibit plateaus (Fig. S2A). The corresponding phase velocities are denoted by $c_\infty^{A0}$ and $c_\infty^{S0}$, respectively. $c_\infty^{A0}$ is equivalent to the speed of the interfacial wave at the fluid-cornea interface (Scholte wave). $c_\infty^{S0}$ is equivalent to the speed of the surface wave at the free surface of the cornea (Rayleigh wave). $c_S = \sqrt{\alpha/\rho}$ is the bulk shear wave speed along the $x_1$ axis. In the absence of prestress (i.e., $\lambda = 1$), $c_S = \sqrt{\mu/\rho}$. The dependences of $c_\infty^{A0}/c_S$ and $c_\infty^{S0}/c_S$ on the material anisotropy ($2k_1/\mu$) are shown in Fig. S2B. $c_\infty^{A0}/c_S$ and $c_\infty^{S0}/c_S$ increase slightly with $2k_1/\mu$ but are always smaller than 1. For isotropic material ($k_1/\mu = 0$), we have $c_\infty^{A0}/c_S \approx 0.839$, $c_\infty^{S0}/c_S \approx 0.955$. These values close to 1 indicate that $c_\infty^{A0}$ and $c_\infty^{S0}$ are close to the bulk shear wave speed $c_S$, and therefore probing $c_\infty^{A0}$ and $c_\infty^{S0}$ gives us a good estimate of the shear modulus $\mu$.

The phase velocity of S0 at zero frequency is nonzero and denoted by $c_0^{S0}$. It can be shown that $c_0^{S0} = \sqrt{(2\beta + 2\gamma)/\rho}$. In the absence of prestress (see Eq. (S10)), we have $c_0^{S0} = 2\sqrt{(\mu + k_1)/\rho}$ (Fig. S2B), from which we find that probing $c_0^{S0}$ gives us the access to the anisotropic material parameter $k_1$. In fact, $c_0^{S0}$ is determined by the plane-strain Young's modulus along the $x_1$ axis, which is the ratio between the uniaxial stress and the strain when the material is stretch along the $x_1$ axis while the out-of-plane direction ($x_3$) is completely constrained. We find that the plane-strain modulus $E_1^*$ is given by $E_1^* = 4\mu + 4k_1$. For isotropic materials ($k_1/\mu = 0$), $E_1^*$ is four times the shear modulus, a well-known result for incompressible materials (Poisson's ratio is 0.5).

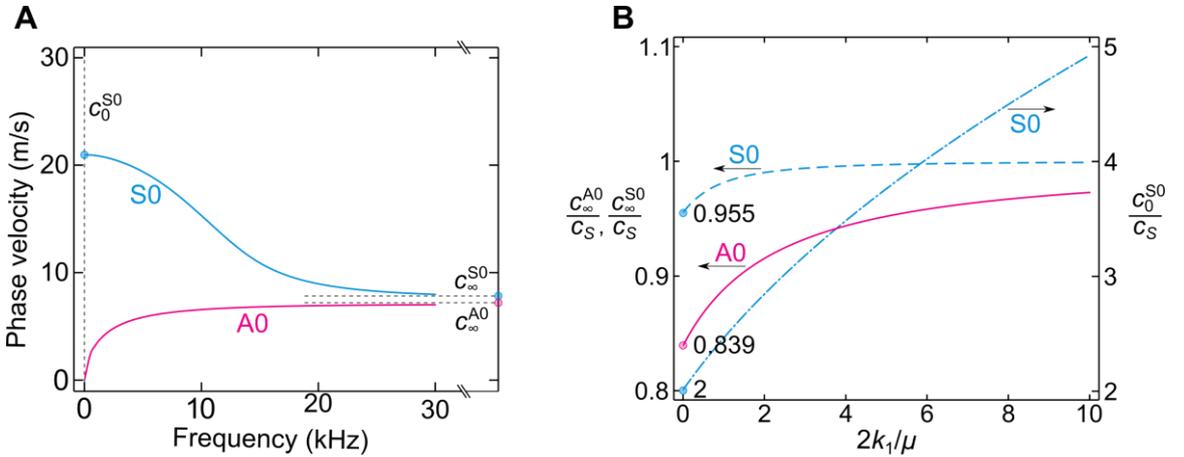

**Fig. S2** (A) Phase velocities of A0 and S0 waves over frequency. $c_\infty^{A0}$ and $c_\infty^{S0}$ denote the phase velocities of the A0 and S0 waves at infinite frequency. $c_0^{S0}$, phase velocity of the S0 at zero frequency. (B) Dependences of $c_\infty^{A0}$, $c_\infty^{S0}$, and $c_0^{S0}$ on the material anisotropy. For isotropic materials ($2k_1/\mu = 0$), we obtain $c_\infty^{A0}/c_S \approx 0.839$ (Scholte wave), $c_\infty^{S0}/c_S \approx 0.955$ (Rayleigh wave), and $c_0^{S0}/c_T = 2$.

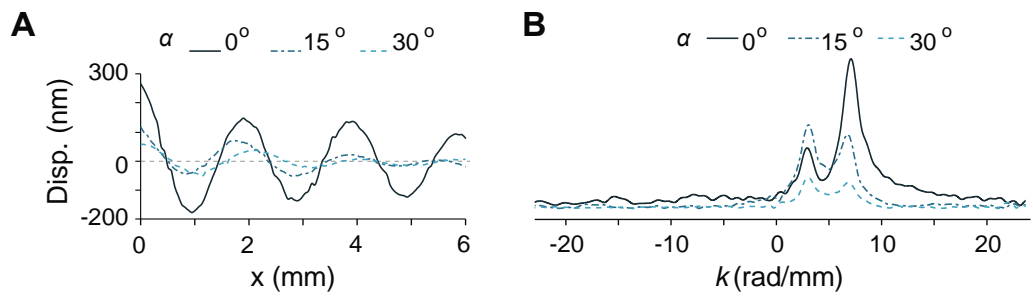

**Fig. S3** Optimization of the probe tip angle $\alpha$. (A) Total displacement amplitude at 4 kHz at different angles. (B) Wavenumber domain plot showing the amplitude ratio of the S0 and A0 waves at 12 kHz.

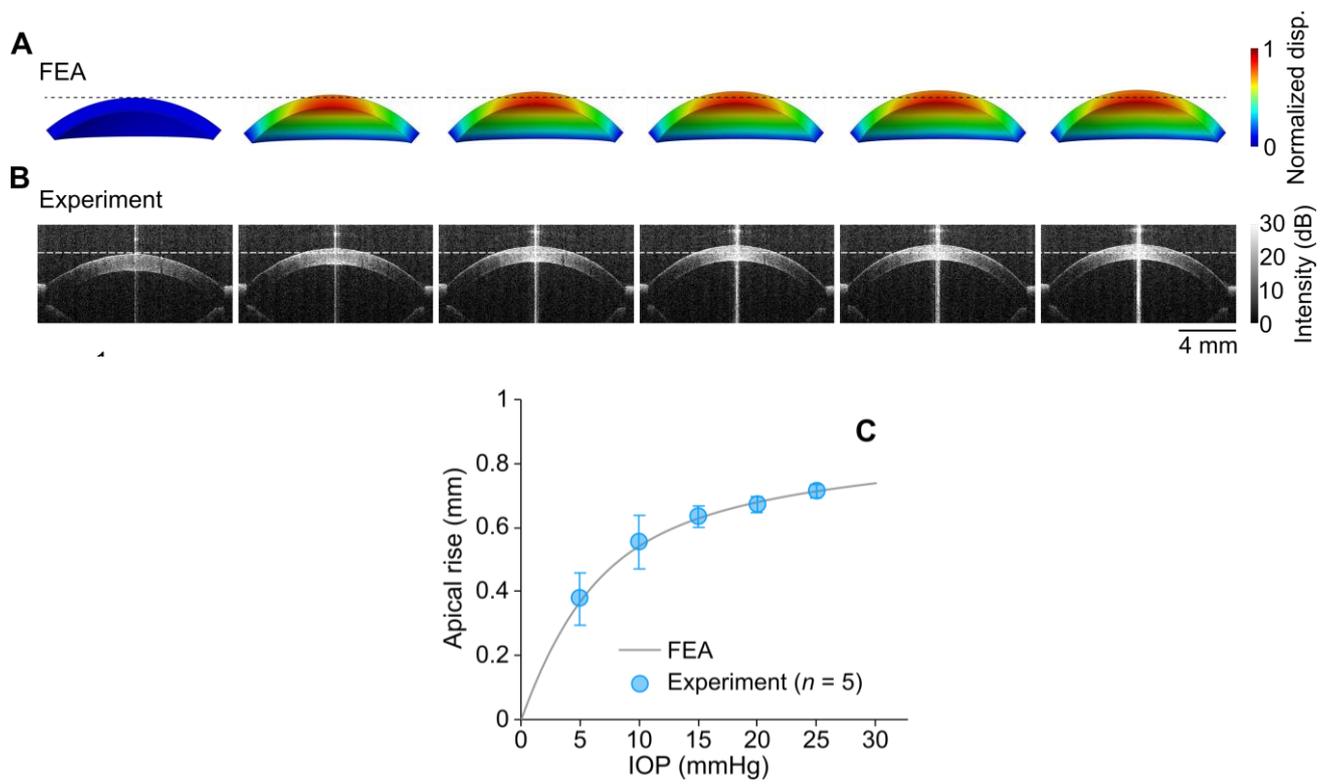

**Fig. S4** Inflation test of the porcine cornea. (A) FEA to show deformation of the cornea. (B) OCT images of the inflated corneas as the intraocular pressure is increased. From left to right, IOP = 0, 5, 10, 15, 20, and 25 mmHg. (C) Comparison of the apical rise between FEA and experiment (N=5).